\documentclass[pre,aps,twocolumn]{revtex4}

\usepackage{graphicx}
\usepackage{amsmath}
\usepackage{amssymb}
\usepackage{ifthen}
\usepackage{booktabs}

\usepackage{graphicx}
\usepackage{dcolumn}
\usepackage{bm}
\usepackage{longtable}

\newcommand{\bb}{\begin{equation}}
\newcommand{\ee}{\end{equation}}
\newcommand{\ba}{\begin{eqnarray*}}
\newcommand{\ea}{\end{eqnarray*}}
\newcommand{\rhor}{\rho({\bf r})}
\newcommand{\dd}{{\rm d}}
\newcommand{\rr}{{\mathbf r}}
\newcommand{\dr}{{\rm d}{\bf r}}

\begin{document}

\title{Filling and wetting transitions at grooved substrates}

\author{Alexandr \surname{Malijevsk\'y}}
\affiliation{
{E. H{\'a}la Laboratory of Thermodynamics, Institute of Chemical Process Fundamentals, Academy of Sciences, 16502 Prague 6, Czech Republic}\\
{Department of Physical Chemistry, Institute of Chemical Technology, Prague, 166 28 Praha 6, Czech Republic}}

\begin{abstract}
The wetting and filling properties of a fluid adsorbed on a solid grooved substrate are studied by means of a microscopic density functional theory. The grooved substrates are modelled using a solid slab, interacting with the
fluid particles via long-range dispersion forces, to which a one-dimensional array of infinitely long rectangular grooves is sculpted. By investigating the effect of the groove periodicity and the width of the grooves and the
ridges, a rich variety of different wetting morphologies is found. In particular, we show that for a saturated ambient gas, the adsorbent can occur in one of four wetting states characterised by i) empty grooves, ii) filled
grooves, iii) a formation of mesoscopic hemispherical caps iv) a macroscopically wet surface. The character of the transition between particular regimes, that also extend off-coexistence, sensitively depends on the model
geometry. A temperature at which the system becomes completely wet is considerably higher than that for a flat wall.
\end{abstract}

\maketitle

\section{Introduction}

Wetting and related phenomena at planar surfaces have been thoroughly studied for the last several decades and are currently fairly well  understood at least for simple fluids \cite{dietrich, sullivan, schick}. More recently,
the subject of fluid adsorption on structured surfaces has received considerable attention from researchers and engineers \cite{quere}. From an engineering viewpoint, the perspective on materials that possess large
surface-to-volume ratios has been appreciated. Indeed, recent advances in lithography have allowed the decoration of solid surfaces on the micro- and nano-scales and facilitated the fabrication of such devices
\cite{whitesides}.
From a more fundamental viewpoint, it was recognised that structured substrates, when exposed to a gas that is close to coexistence with its liquid phase, can
produce quite distinct adsorption characteristics compared to planar systems \cite{cacamo, nature}. For instance, macroscopic considerations that are based on
Young's and Laplace's equations predict that the adsorption properties of a substrate with a linear-wedge shape can be sensitively controlled by its opening
angle \cite{hauge}. As a consequence, a large amount of the emerged liquid phase may adsorb near the apex even though only a microscopic film of the liquid is
adsorbed far from the wedge apex. However, statistical mechanics has to be incorporated to learn more about the nature of interfacial phenomena on non-planar
surfaces, particularly with respect to the phase transitions that they induce. Among other findings, such a more microscopic approach has resulted in the
discovery of interesting hidden symmetries (or so-called covariances) that relate the adsorption properties of different substrate geometries \cite{cov, tasin}.

\begin{figure}[h]
\includegraphics[width=7.5cm]{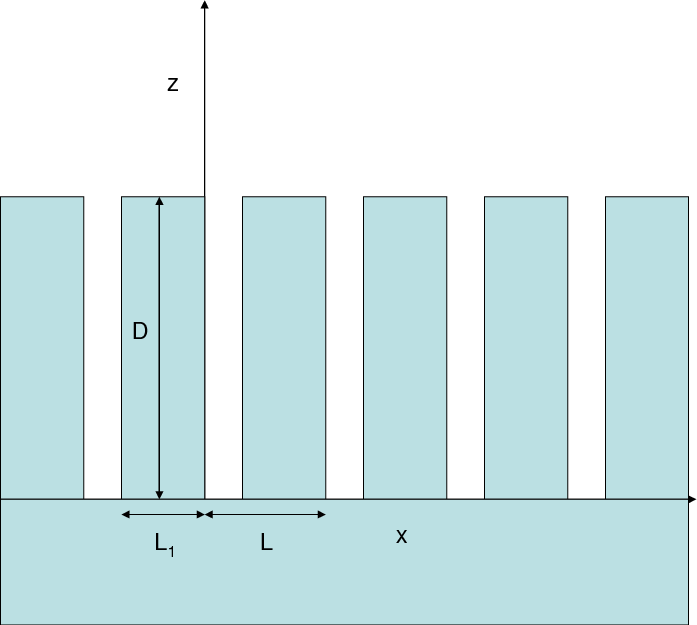}
\caption{Illustration of the model. The grooves (each of width $L_2=L-L_1$ and depth $D$) form an infinite one-dimensional array along the $x$-axis. The sketch
corresponds to the projection $y={\rm const}$.}\label{sketch}
\end{figure}

One class of structured surfaces that has recently attracted a strong interest motivated by recent experiments \cite{bruschi1, bruschi2, javadi} comprises solid
substrates patterned with regular arrays. A sketch of such a model substrate used in this work is provided in Fig.~\ref{sketch}. This ``grooved substrate'' is
characterised by the presence of rectangular capillary grooves of depth $D$ and width $L_2$, which are etched into a solid slab. The grooves form an infinite
periodic one-dimensional array along the $x$-direction with a periodicity $L$ and are assumed to be unbounded in the $y$-direction. It is further assumed that
the substrate is formed with uniformly distributed atoms, which interact with the fluid particles with a long-range potential that decays as $1/r^6$ at large
distances.


Related substrate models were considered rather recently by Dietrich {\it et al.} \cite{tasin, hofmann, checco} to study complete wetting of geometrically
structured substrates by interfacial Hamiltonian theory, derived from the so-called sharp-kink approximation of density functional theory (DFT) \cite{dietrich}.
In this paper more microscopic model is used, which takes into account the short-range correlations between fluid particles by adopting Rosenfeld's fundamental
measure theory (FMT) \cite{ros}. In particular, this allows to address the questions such as:  What is the equilibrium structure of an adsorbed fluid for a
particular substrate geometry? How does the structure respond to a change in temperature and bulk pressure (chemical potential)? What type of phase transitions
does the system exhibit and how does this depend on the individual parameters that characterise the substrate geometry?

The remainder of the paper is organised as follows. In Section II, fundamental adsorption phenomena at substrates, to which the current model reduces in special cases, is briefly recalled. In Section III, we formulate the
molecular model of the substrate and the fluid and outline the main features of the microscopic DFT used in this work. The numerical results including surface phase diagrams for several representative geometrical models are
presented in Section IV. The main points of the work are summarised and discussed in Section V and a link with the earlier works based on interface Hamiltonian is made. The paper concludes with the appendices, which provide
the details of the derivations of some formulae presented earlier in the paper.

\section{Behaviour of the model substrate in special cases}

In this section, we make a link between the substrate model sketched in Fig.~\ref{sketch} with more familiar model substrates to which the current model reduces
after taking the special limits of the substrate geometric parameters. The main adsorption characteristics of the resulting systems are briefly recalled.

\subsection{$L_1\to\infty$ or $L_2\to\infty$ or $D\to0$}

If any of these limits is realised (alternatively, one can also take $L_1\to0$ or $L_2\to0$), the grooved substrate reduces to a simple planar wall. At the liquid-vapour bulk coexistence, the planar wall (preferentially
adsorbing liquid phase) exhibits wetting transition at the wetting temperature $T_w$. Typically, the transition is first-order but can also be continuous (critical) depending on the range and strength of the inter-molecular
interactions. It should be noted that the latter phenomenon is notably rare in nature; in fact it has not been observed for solid substrates and has only been detected for a few binary liquid mixtures.

Alternatively, the wetting layer may develop at a constant temperature $T>T_w$ upon approaching the saturation value of the chemical potential $\mu\to\mu_{\rm
sat}(T)^-$. In this complete wetting process the width of the liquid film $\ell_\pi$ develops according to
 \bb
 \ell_\pi\sim \delta\mu^{-\beta_s}\,,\label{ell_pi}
 \ee
as $\delta\mu=\mu_{\rm sat}(T)-\mu\to0$. The critical exponent  $\beta_{\rm s}=1/3$ if the dominating force at large distances originates from dispersion
interactions. If the wetting transition at $T_w$ is first order, the singularity of the first derivative of the free energy at $T_w$ is prolonged off-coexistence
in a pre-wetting line representing the loci of the thin-thick first order transitions. This line terminates at its own critical temperature $T_{\rm sc}$ and
approaches the coexistence line tangentially at $T_w$  as $|\delta\mu|\sim(T-T_w)^{3/2}$.

\subsection{$L\to\infty$ or $D\to\infty$}

Either of the two limits defines a capped capillary (single groove). Because of the presence of the bottom wall, a meniscus separating capillary-gas and capillary-liquid is formed near the bottom for $\mu<\mu_{cc}(H)$, whereas
for $\mu>\mu_{cc}(H)$ the pore must be filled with a liquid, so that the meniscus is to be found at the top. Here, $\mu_{cc}(H)$ is the chemical potential of a slit (parallel plate) pore of a width $H$, at which the capillary
condensation occurs. Interestingly, in contrast with the capillary condensation in a slit pore, which is a first-order transition, the condensation in the capped capillary prove to be continuous for walls that are completely
wet \cite{evans_cc, marconi, darbellay, tasin, gelb, roth, malijevsky_cc, parry_03}. The condensation is then given by an unbinding of the meniscus separating the capillary-gas and the capillary-liquid from the bottom end
according to the power law \cite{evans_cc}
 \bb
 \ell_{cc}\sim (\mu-\mu_{cc}(H)^{-\beta_{cc}}\,,\label{ell_cc}
 \ee
with $\beta_{cc}=\frac{1}{4}$ for long range forces. In the most recent studies it was revealed \cite{malijevsky_cc, parry_03} that the order of the transition
is controlled by the wetting regime of the bottom wall, such that below $T_w$ the adsorption in a capped capillary exhibits first-order transition.

While these features are experienced by single grooves specified by either of the two limits, it should be noted that the phase transitions and criticality of
the former one with $D$ finite must be necessarily rounded beyond the mean-filed approximation (1D Ising model universality class).

\subsection{$D\to\infty$ and $L_2\to\infty$}

In this case, the grooved substrate reduces to a geometry of a linear wedge with an opening angle $\psi=\pi/2$. Such a model exhibits wedge filling transition
that differs from both wetting and capillary condensation: at the vapour-liquid coexistence, a wedge becomes completely filled with liquid at the filling
temperature $T_f$ that corresponds to the contact angle of the liquid drop on the wall \cite{hauge, rejmer, wood1}:
 \bb
 \theta(T=T_f)=(\pi-\psi)/2\,.
 \ee
The filling transition is critical whenever the corresponding wetting transition is also critical; however, if the wetting transition is first-order, the filling
transition can be either first-order or critical depending on the opening angle \cite{milchev, malijevsky_prl, malijevsky_wedge}.

\section{Theory}

\subsection{Potential of the grooved substrate}\label{wall}

The model grooved substrate, as sketched in Fig.~\ref{sketch}, enters into the theory as an external field $V(\rr)$ exerted on the fluid atoms. Owing to the periodicity along the $x$-axis and its translational invariance in
the $y$-dimension, one can write :
 \bb
 V(x,z)=V_1(z)+\sum_{n=-\infty}^\infty V_2(x+nL,z)\,. \label{pot} 
 \ee
Here, $V_1(z)$ is the potential of a planar wall $\mathbb{W}=\{x\in\mathbb{R}, y\in\mathbb{R}, z\in(-\infty,0)\}$, and $V_2(x,z)$ is the potential of a rectangular body $\mathbb{B}=\{x\in(-L_1,0),y\in\mathbb{R},z\in(0,D)\}$
with a property $V_2(x,z)=V_2(-x-L_1,z)$.

The substrate is treated as a continuous distribution of atoms with a one-body density $\rho_w$, each of which interacts with the fluid atoms via a Lennard-Jones tail
 \bb
 \phi_w(r)=-4\varepsilon_w\left(\frac{\sigma}{r}\right)^{6}\,, \label{phiw}
 \ee
where $r$ is the distance between the substrate and the fluid particles. After integrating $\phi_w(\rr)$ over the entire domain of the wall (see Appendix A) and introducing a hard-wall barrier to model a short-range repulsion
between fluid and wall atoms, the potential of the substrate can be expressed as follows:

 \bb
 V_1(z)=\left\{\begin{array}{ll} \tilde{V_1}(z)\,; & z>\sigma\,,\\
\infty\,;& z<\sigma\end{array}\right. \label{V1}
 \ee

 and

 \bb
 V_2(x,z)=\left\{\begin{array}{ll} \infty\,; & x\in(-L_1-\sigma,\sigma)\cap z\in(0,D+\sigma)\,,\\
 \tilde{V_2}(x,z)\,; &{\rm otherwise\,,}\end{array}\right. \label{V2}
 \ee

with

\bb
 \tilde{V_1}(z)=\frac{2\alpha_w}{z^3}\,,\label{V1_att}
\ee

\bb
 \tilde{V_2}(x,z)=\alpha_w\left[\psi_{z,D}(L_1+x)-\psi_{z,D}(x)\right]\,, \label{V2_att}
\ee

\bb
 \alpha_w\equiv-\frac{1}{3}\pi\varepsilon_w\rho_w\sigma^6 \label{alpha_w}
\ee

and

\begin{eqnarray}
 \psi_{z,D}(x)&\equiv& \frac{2x^4+x^2(z-D)^2+2(z-D)^4}{2x^3(z-D)^3\sqrt{x^2+(z-D)^2}}\nonumber\\
 &&-\frac{2x^4+x^2z^2+2z^4}{2x^3z^3\sqrt{x^2+z^2}} \,.\label{psi}
\end{eqnarray}

\subsection{Density functional theory}


Within classical density functional theory \cite{evans79}, the equilibrium density profile is obtained by minimising the grand potential functional
 \bb
 \Omega[\rho]={\cal F}[\rho]+\int\dd\rr\rhor[V(\rr)-\mu]\,,\label{om}
 \ee
where $\mu$ is the chemical potential and $V(\rr)$ is the external potential. Here, ${\cal F}[\rho]$ is the intrinsic free energy functional of the fluid one-body density, $\rhor$, which can be split into ideal and excess
parts. As is common in modern DFT, the excess free energy functional is further divided into a hard-sphere term and an attractive contribution
  \bb
  {\cal F}_{\rm ex}[\rho]={\cal F}_{\rm hs}[\rho]+\frac{1}{2}\int\int\dd\rr\dd\rr'\rhor\rho(\rr')u_{\rm a}(|\rr-\rr'|)\,, \label{f}
  \ee
where  $u_{\rm a}(r)$ is the attractive portion of the fluid-fluid interaction potential. In our analysis, we consider this to be a truncated Lennard-Jones-like
potential
 \bb
 u_{\rm a}(r)=\left\{\begin{array}{cc}
 0\,;&r<\sigma\,,\\
-4\varepsilon\left(\frac{\sigma}{r}\right)^6\,;& \sigma<r<r_c\,,\\
0\,;&r>r_c\,.
\end{array}\right.\label{ua}
 \ee
which is cut-off at $r_c=2.5\,\sigma$, where $\sigma$ is the hard-sphere diameter. It should be noted that since the range of the wall-fluid and fluid-fluid
interactions are different, the Hamaker constant of the system is always positive (at least within the sharp-kink approximation), which ensures that wetting
transition at the corresponding planar wall is first order \cite{dietrich}. Note that the same molecular model was used recently in Refs.\, \cite{malijevsky_prl,
malijevsky_wedge}.

The hard-sphere part of the excess free energy is approximated by the FMT functional \cite{ros},
 \bb
{\cal F}_{\rm hs}[\rho]=\frac{1}{\beta}\int\dd\rr\,\Phi(\{n_\alpha\})\,,
 \ee
where  $\beta=1/k_BT$ is the inverse temperature. The function $\Phi$ depends on six weighted densities $n_\alpha(\rr)$ which can be expressed as double integrals for the rectangular symmetry as is dictated by the external
field (\ref{pot}) (see Appendix B).

The minimisation of (\ref{om}) results in an Euler-Lagrange equation
 \bb
 \mu_{\rm id}(\rhor)+\frac{\delta{\cal F}_{\rm hs}[\rho]}{\delta\rho(\rr)}+\int\dd\rr'\rho(\rr')u_{\rm a}(|\rr-\rr'|)=\mu\,,\label{el}
 \ee
where $\mu_{\rm id}$ is the ideal part of the chemical potential. The equilibrium density profile is obtained by solving  Eq.~(\ref{el}) numerically on a two-dimensional Cartesian grid with a spacing of $0.1\,\sigma$. The
convolution term in (\ref{el}) is recast into a two-dimensional integral (see Appendix C).

The thermodynamic properties of the corresponding bulk system are determined from Eq.~(\ref{om}), which is applied to $\rhor=const$. From the resulting Helmholtz free energy the liquid-vapour phase boundary can be constructed.
The binodal terminates at the bulk critical temperature $k_BT_c/\varepsilon=1.41$.

\section{Numerical results}

Throughout this paper, we will adopt $\sigma$ and $\varepsilon$ as our length and energy units. Therefore, we will express our quantities in dimensionless units,  such as  $T^*=k_BT/\varepsilon$, $\rho^*=\rho\sigma^3$,
$x^*=x/\sigma$, etc., unless stated otherwise. The interaction potential that characterise the substrate strength is set to $\varepsilon_w=1.2\,\varepsilon$,  and we will restrict ourselves to the results for a substrate depth
of $D=5\sigma$.

\begin{figure}[h]
\includegraphics[width=0.4\textwidth]{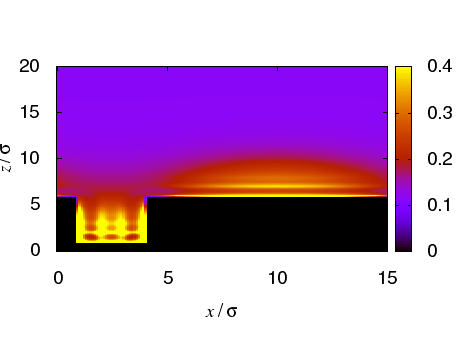}
\includegraphics[width=0.4\textwidth]{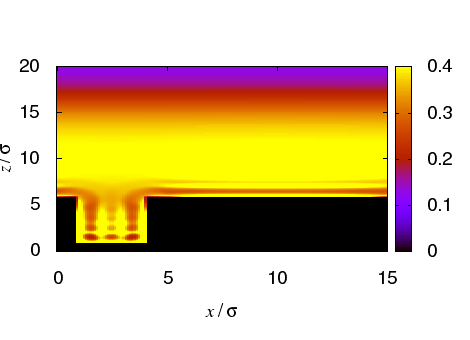}
\caption{Coexisting density profiles for a model  with $L^*=15$, $L_1^*=10$, and $D^*=5$. The temperature of the systems is $T^*_{gw}=1.33$ and the bulk phase is a saturated gas. }\label{pD5L15L1=10}
\end{figure}

\begin{figure}
\includegraphics[width=0.44\textwidth]{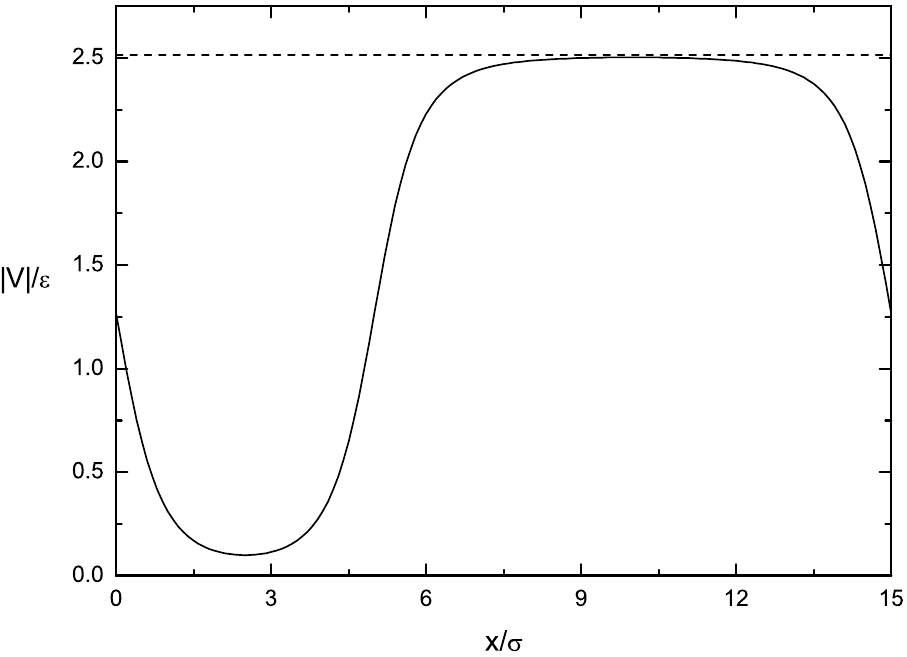}
\caption{Absolute value of the potential of the grooved-surface model with $L^*=15$, $L_1^*=10$, and $D^*=5$ at $z=D+\sigma$ (solid line) and the potential of the planar wall at $z=\sigma$ (dashed line). }\label{potential}
\end{figure}

\subsection{Coexistence path}

All results presented in this paragraph are for the liquid-gas coexistence with a boundary condition for the density profile $\rho(x,z_f)=\rho_v(T)$ for all $x$, where $\rho_v(T)$ is the particle density of a saturated vapour
and $z_f$ refers to a vertical size of the system. Our primary goal is to inspect the type of wetting regimes that a system with a particular geometry can acquire and compare with the wetting properties of the corresponding
planar wall.
When applied to a planar wall, the solution of the Euler-Lagrange equation (Eq.~(\ref{el})) predicts a first-order wetting transition at a temperature $T_w^*=1.17$ \cite{malijevsky_prl}.



\begin{figure}[h]
\includegraphics[width=0.4\textwidth]{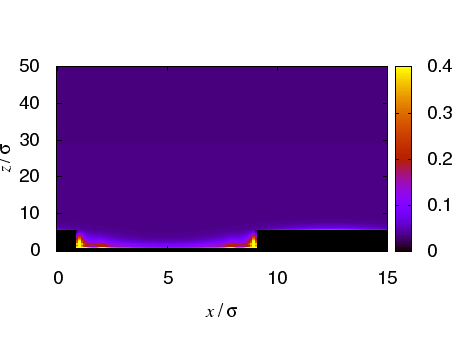}
\includegraphics[width=0.4\textwidth]{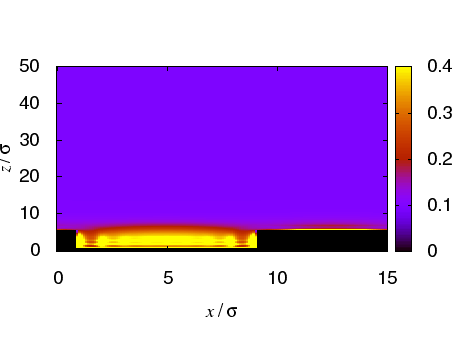}
\includegraphics[width=0.4\textwidth]{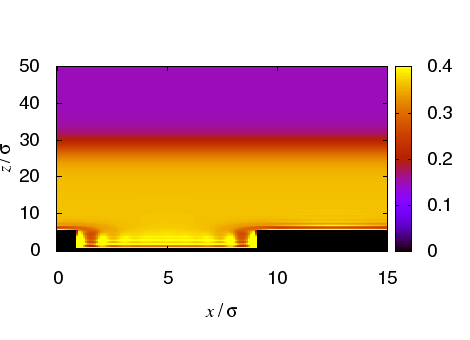}
\caption{Density profiles of three distinct (separated by first-order transitions) phases for a model with $L^*=15$, $L_1^*=5$, and $D^*=5$. The corresponding temperatures are (from above): $T^*=1.1$ , $T^*=1.3$ and
$T^*=1.37$. In all cases the bulk phase is the saturated gas.}\label{pD5L15L1=5}
\end{figure}

\begin{figure}[h]
\includegraphics[width=0.4\textwidth]{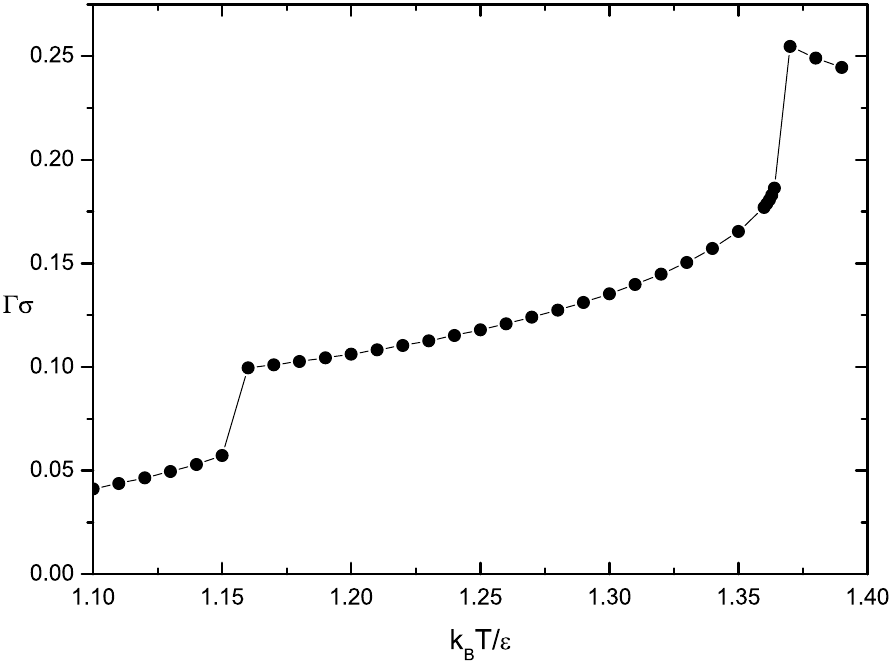}
\caption{Temperature dependence of the adsorption on a bulk coexistence line for a model with $L^*=15$, $L_1^*=5$, and $D^*=5$. }\label{D5L15L1=5}
\end{figure}

\begin{figure}[h]
\includegraphics[width=0.4\textwidth]{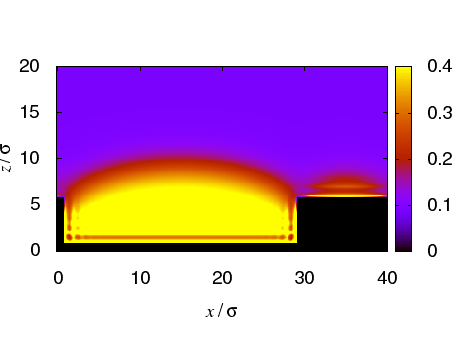}
\includegraphics[width=0.4\textwidth]{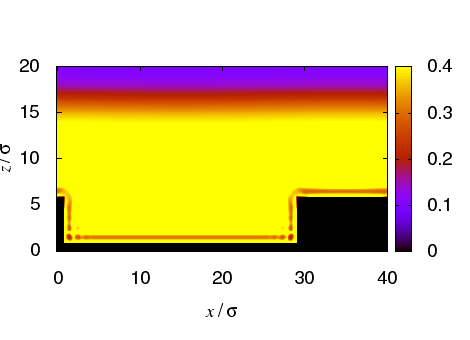}
\caption{Coexisting density profiles for a model with $L^*=40$, $L_1^*=10$, and $D^*=5$. The temperature of the systems is $T^*_{gw}=1.19$  and the bulk phase is the saturated gas. }\label{pL40L1=10}
\end{figure}

We start by considering a substrate model characterised by the parameters $L^*=15$ and $L_1^*=10$. In this case, the minimisation of the grand potential functional (\ref{om}) leads to two different sets of results depending on
the initial configuration. The temperature, at which two different states yield the identical grand potential value, defines the location of the first order transition, and we will refer to this process as
\emph{groove-wetting}. This process is now compared with an ordinary wetting transition on a planar wall. First, as observed from the plots in  Fig.~\ref{pD5L15L1=10} in which the coexisting low- and high-density profiles are
displayed (for a single period), the symmetry breaking in the $x$-dimension induces a non-planar character of the liquid-vapour interface in the low-density state. This behaviour is in contrast with a liquid-vapour interface
above a flat wall, in which case only the fluctuation effects disrupt its otherwise planar shape. The high-density state corresponds to a completely wet substrate, so that the non-uniformity of the substrate potential in the
$x$-direction no longer influences the geometry of the unbounded liquid-gas interface. Second, the groove-wetting occurs at a temperature $T_{gw}^*=1.33$, which is much closer to the bulk critical temperature (recall,
$T_c^*=1.41$) than the wetting temperature of a planar wall, $T_w^*=1.17$. This result can be explained by comparing the strength of the potentials of the flat substrate and that of the grooved substrate. As shown in
Fig.~\ref{potential}, the latter reaches the value of the planar wall about the middle of the ridges but the presence of the grooves significantly lowers the local absolute value of the potential at a given height above the
substrate. Thus, the grooved substrate constitutes an effectively weaker adsorbent than the corresponding planar wall, which pushes the temperature at which the surface is completely wet upwards.


For thicker grooves, $L_1^*=5$ (while maintaining $L^*=15$), the adsorption scenario changes as the system can now realise three different regimes as displayed in Fig.~\ref{pD5L15L1=5}. These regimes are separated by two
first-order transitions -- see Fig.~\ref{D5L15L1=5}, where the temperature dependence of the adsorption per unit length is shown. The latter is defined as
  \bb
 \Gamma=\int\int\dd x \dd z(\rho(x,z)-\rho_b)\,,\label{ads}
 \ee
where in the current case referring to a bulk coexistence state $\rho_b=\rho_v(T)$, and the integral is taken over the volume of the system that is available for fluid particles within an interval $x\in(0,L)$. In contrast with
the previous case, the width of the grooves is now sufficient to allow \emph{groove-filling} first-order phase transition before the system experiences groove-wetting. Beyond this, the system is already saturated, and the
adsorption is effectively infinite.

\begin{figure}[h]
\includegraphics[width=0.4\textwidth]{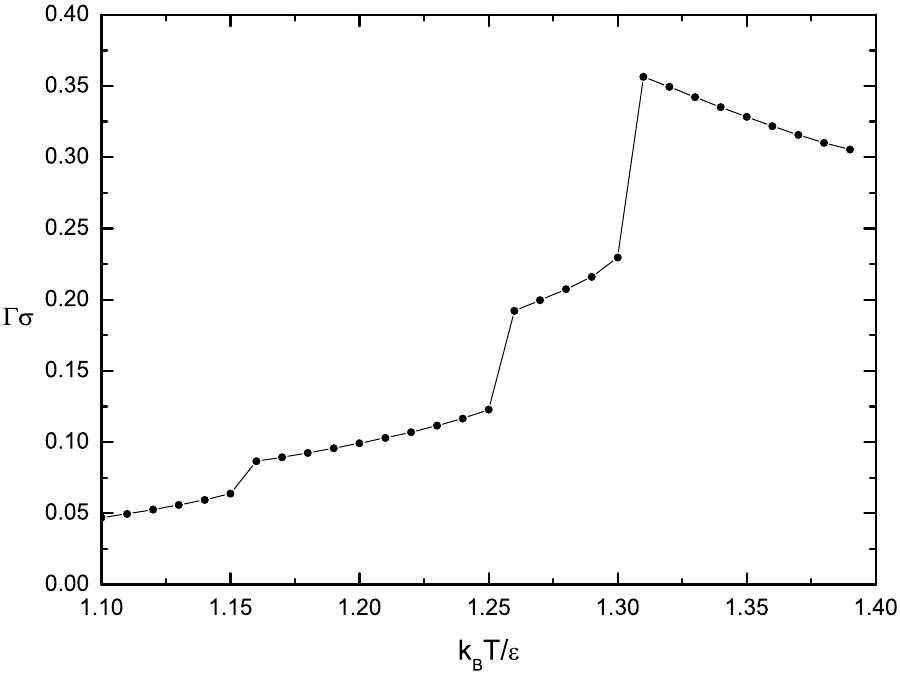}
\caption{Temperature dependence of the adsorption on a bulk coexistence line for a model with $L^*=40$, $L_1^*=30$, and $D^*=5$. For $T^*>1.3$, the system is already saturated (because of its finite size). }\label{L40L1=30}
\end{figure}

\begin{figure}[h]
\includegraphics[width=0.4\textwidth]{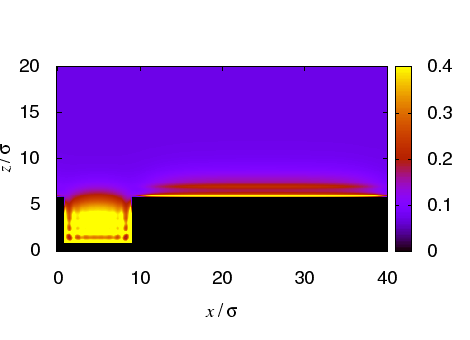}
\includegraphics[width=0.4\textwidth]{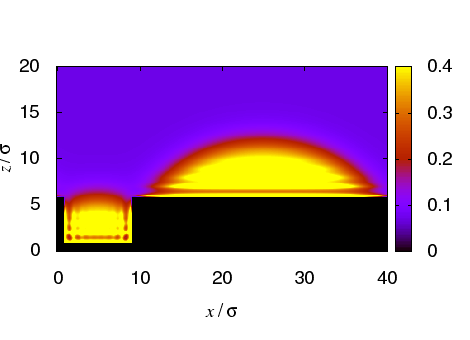}
\caption{Coexisting density profiles for a model with $L^*=40$, $L_1^*=30$, and $D^*=5$. The temperature of the systems is $T^*_{gw}=1.24$ and the bulk phase is the saturated gas. }\label{pL40L1=30}
\end{figure}

We now increase the periodicity to $L^*=40$ and examine the impact of the groove width on fluid adsorption. For $L_1^*=10$, the situation is similar to the observations from the previous case: the system undergoes
groove-filling ($T_{gf}^*=1.16$) and groove-wetting transitions ($T_{gw}^*=1.29$). Compared to the case where $L^*=15$, the wall inhomogeneity is less dramatic and the complete wetting of the wall takes place closer to $T_w$.
Now, below, but near $T_{gw}$, separate wetting layers with concave interfaces are formed at the grooves and the ridges, and groove-wetting occurs when the two layers merge, such that the resulting liquid-vapour interface
becomes flat, cf. Fig.~\ref{pL40L1=10}. When the groove width decreases (increasing $L_1$), both $T_{gf}$ and $T_{gw}$ progressively decrease and for a sufficiently large $L_1$ a new first-order transition is revealed as
illustrated in Fig.~\ref{L40L1=30} for $L_1^*=30$. In addition to groove-filling ($T_{gf}=1.09$) and groove-wetting ($T_{gw}=1.23$), there is also an equilibrium between two density profiles that differ by the amount of the
adsorbed liquid at the ridges as displayed Fig.~\ref{pL40L1=30}.  A sufficiently large value of $L_1$ enables the abrupt development of a thick layer at the ridges as a result of the partial unbinding of the liquid-vapour
interface being pinned at the edges. We will refer to this process as \emph{bounded-wetting}. The temperature of this transition $T_{bw}$ also decreases with $L_2$, and the transition is metastable with respect to
groove-wetting unless $T_{bw}<T_{gw}$. For the periodicity of $L^*=40$, this relationship holds for $L_1^*\lesssim33$.


\subsection{Surface phase diagrams}\label{spd}

Now we extend our considerations to off-coexistence states. In this case, the minimisation of the grand potential (\ref{om}) is subject to the boundary condition $\rho(z)=\rho_b$, where $\rho_b$ is a bulk density of a gas with
a chemical potential $\mu\leq\mu_{\rm sat}$. At low values of $\mu$, the adsorption (\ref{ads}) is microscopic, while its upper limit value (at $\mu_{\rm sat}$) is known from the previous considerations. We wish to learn the
behaviour of $\Gamma$ between these two limits.

\begin{figure}[h]
\includegraphics[width=0.4\textwidth]{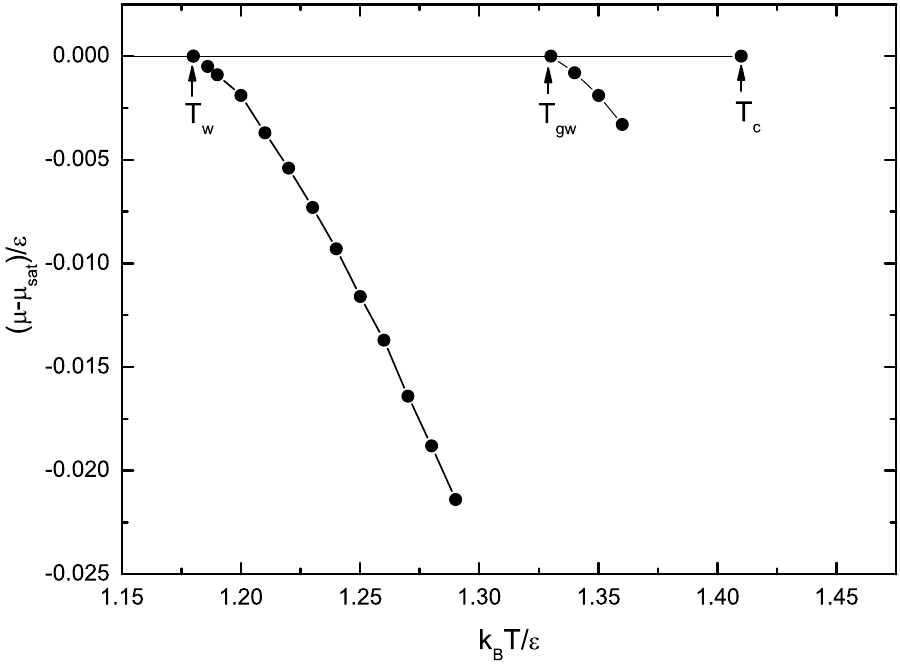}
\caption{Phase diagram for a model with $L^*=15$, $L_1^*=10$, and $D^*=5$. $T_w$ showing the wetting temperature on a flat wall; $T_{gw}$  is the groove-wetting
temperature, and $T_c$ is the bulk critical temperature. For a comparison, the prewetting line of a corresponding flat wall (joining the bulk coexistence at
$T_w$) is also displayed.}\label{pd_L=15_L1=10}
\end{figure}

\begin{figure}[h]
\includegraphics[width=0.4\textwidth]{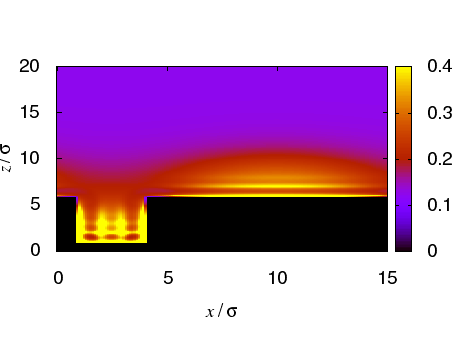}
\includegraphics[width=0.4\textwidth]{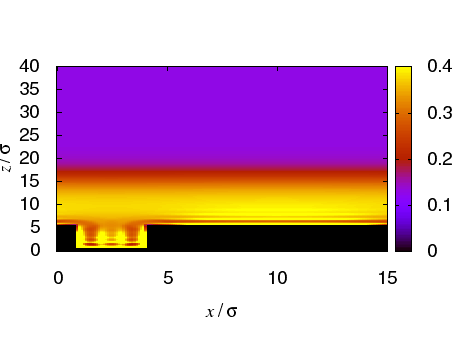}
\caption{Density profiles of the coexisting states before (top panel) and after (bottom panel) groove-prewetting for a  model with $L^*=15$, $L_1^*=10$ and
$D^*=5$. The temperature is $T^*=1.34$.  }\label{p_L=15_L1=10_T=134}
\end{figure}

\begin{figure}[h]
\includegraphics[width=0.4\textwidth]{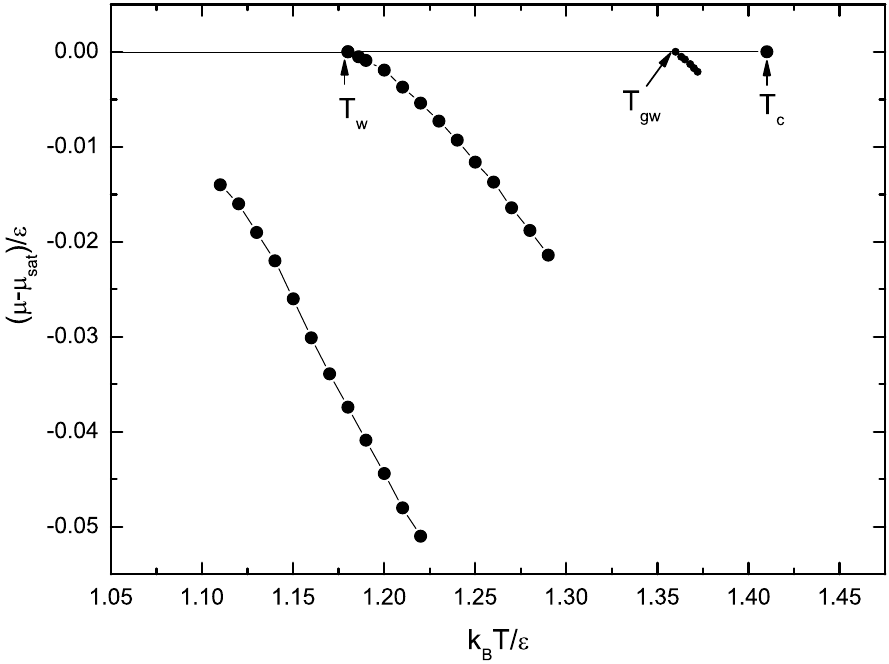}
\caption{Phase diagram for a model with $L^*=15$, $L_1^*=5$, and $D^*=5$.  The nomenclature is same as in Fig.~\ref{pd_L=15_L1=10}. The low temperature curve corresponds to groove-filling, which terminates at its critical
point  at a temperature $k_BT/\varepsilon\approx 1.22$; the left end of the curve, which would normally connect the bulk coexistence (see Fig.~\ref{pd_L=40_L1=20}), is truncated at point, where the first crystal nuclei induced
by the wall occur. For a comparison, the prewetting line of a corresponding flat wall (joining the bulk coexistence at $T_w$) is also displayed.}\label{pd_L=15_L1=5}
\end{figure}

\begin{figure}[h]
\includegraphics[width=0.4\textwidth]{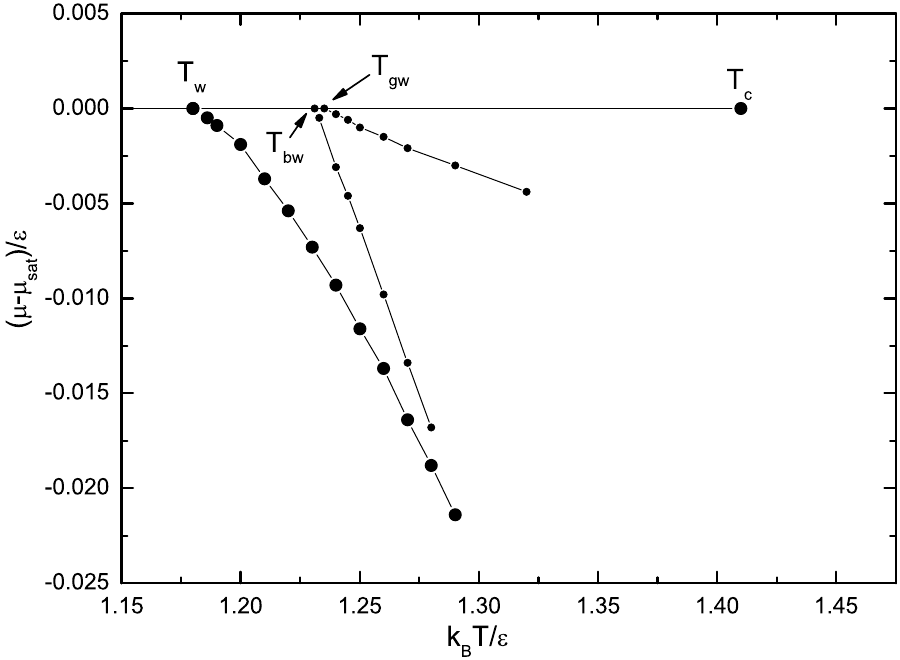}
\caption{Phase diagram for a model with  $L^*=40$, $L_1^*=35$, and $D^*=5$.  The nomenclature is same as in Fig.~\ref{pd_L=15_L1=10}. In addition, the
bounded-wetting temperature $T_{bw}$ is displayed.  For a comparison, the prewetting line of a corresponding flat wall (joining the bulk coexistence at $T_w$) is
also displayed.}\label{pd_L=40_L1=35}
\end{figure}

\begin{figure}[h]
\includegraphics[width=0.4\textwidth]{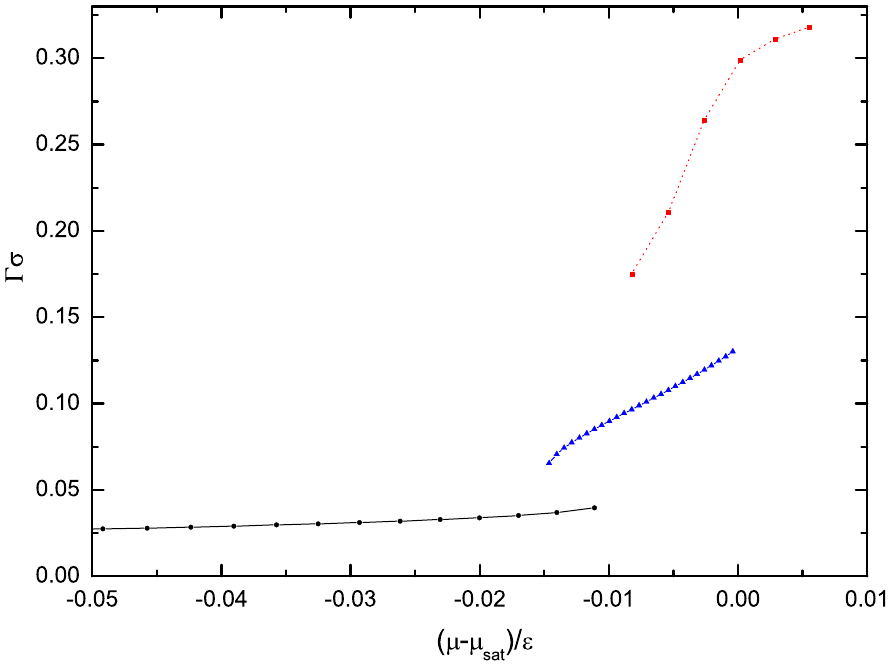}
\includegraphics[width=0.4\textwidth]{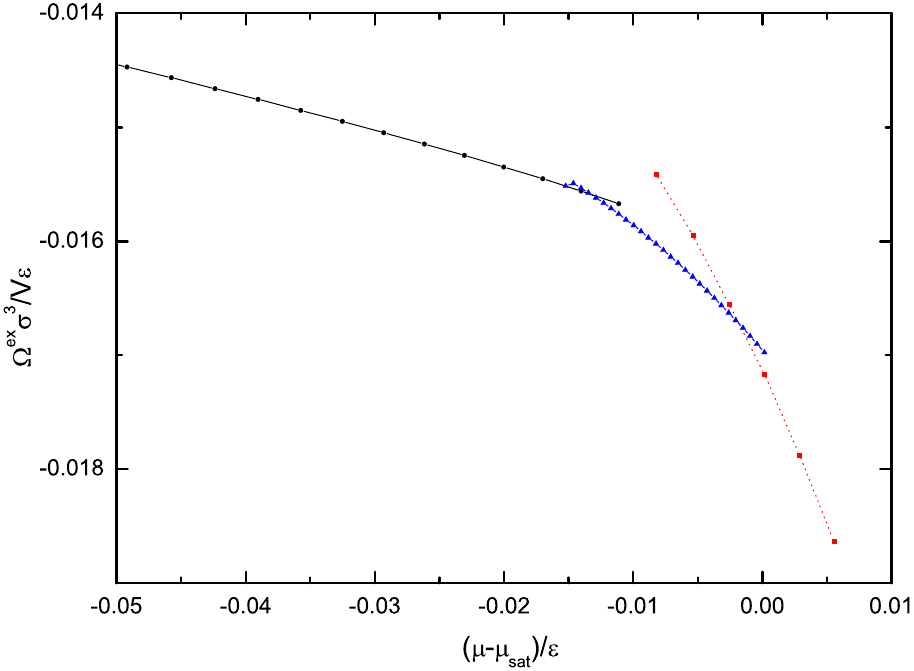}
\caption{Adsorption isotherm (upper panel) and the excess grand potential dependence on chemical potential  for a model with $L^*=40$, $L_1^*=35$, and $D^*=5$ at a temperature $T^*=1.27$.  }\label{L=40_L1=35_T127}
\end{figure}

\begin{figure}[h]
\includegraphics[width=0.4\textwidth]{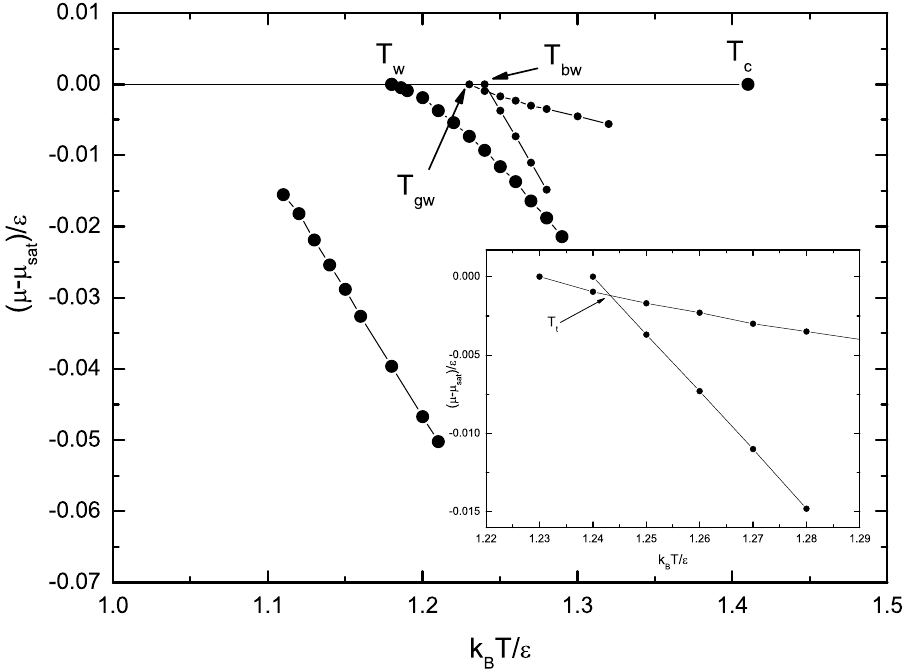}
\caption{Phase diagram for a model with  $L^*=40$, $L_1^*=30$, and $D^*=5$.   The nomenclature is same as in Fig.~\ref{pd_L=40_L1=35}. The symbols at low temperatures correspond to groove-filling. The low temperature curve
corresponds to groove-filling, which terminates at its critical point  at a temperature $k_BT/\varepsilon\approx 1.21$; the left end of the curve, which would normally connect the bulk coexistence (see
Fig.~\ref{pd_L=40_L1=20}), is truncated at point, where the first crystal nuclei induced by the wall occur. For a comparison, the prewetting line of a corresponding flat wall (joining the bulk coexistence at $T_w$) is also
displayed. In the inset, a triple point at which three different wetting morphologies coexist is shown.}\label{pd_L=40_L1=30}
\end{figure}

\begin{figure}[h]
\includegraphics[width=0.47\textwidth]{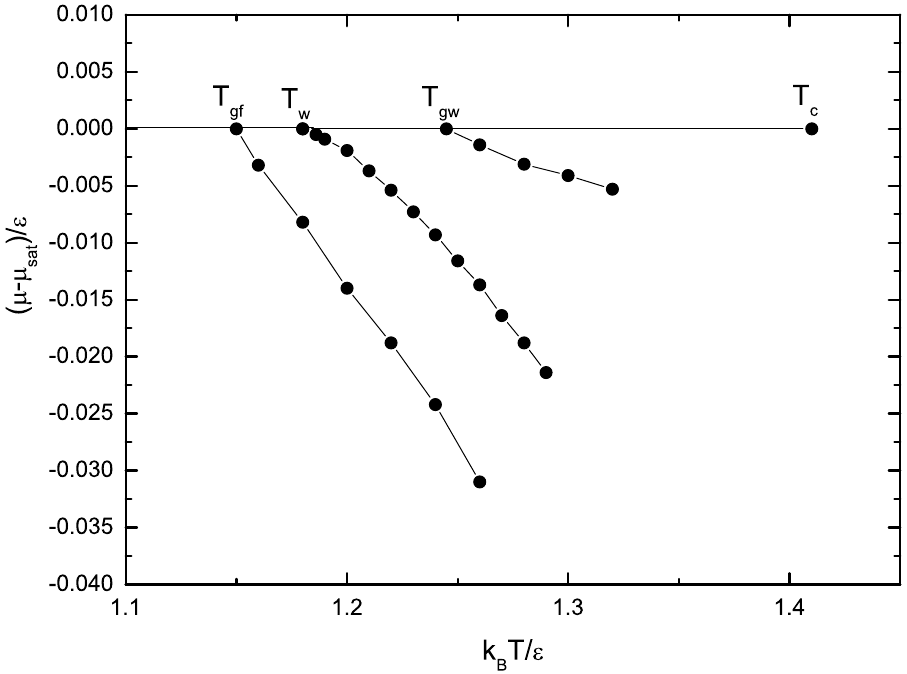}
\caption{Phase diagram for a model with $L^*=40$, $L_1^*=20$, and $D^*=5$. Here, $T_{gf}$ denotes the filling temperature, $T_w$ is the wetting temperature for a planar wall, $T_{gw}$ is the groove-wetting temperature, and
$T_c$ is the bulk critical temperature.  For a comparison, the prewetting line of a corresponding flat wall (joining the bulk coexistence at $T_w$) is also displayed.}\label{pd_L=40_L1=20}
\end{figure}


We start with a periodicity of $L^*=15$. For $L_1^*=10$, the grooves are too narrow  to undergo the filling transition; thus the only free-energy singularity on the coexistence path is at $T_{gw}$, which extends
off-coexistence and terminates at its own critical point as displayed in Fig.~\ref{pd_L=15_L1=10}. The loci of the first-order transitions characterised by a jump in $\Gamma$ are similar to a prewetting line for a thin-thick
transition on a planar wall. Thus, we will call the process \emph{groove-prewetting}. However, in contrast with prewetting on a planar wall, in which case the two coexisting states differ only by a width of an adsorbed film,
here the interface also changes its geometry, as shown in Fig.~\ref{p_L=15_L1=10_T=134}. At the higher-adsorption state (bottom panel in Fig.~\ref{p_L=15_L1=10_T=134}), the fluid inhomogeneity in the $x$-direction has no
substantial influence and the system experiences ordinary complete wetting upon a further increase in $\mu$.

A different scenario occurs for $L_1^*=5$. In this case, the grooves are wide enough that groove-wetting is preceded by the groove-filling transition on the coexistence line. As shown in Fig.~\ref{pd_L=15_L1=5}, the
first-order groove-filling transition also continues off-coexistence and terminates at a temperature $T^*=1.22$. Above this temperature, condensation in the grooves is a continuous process. We further note that below the
temperatures $T^*\lesssim1.1$, groove-filling becomes metastable with respect to a crystalline structure of the fluid atoms induced by the presence of the wall.


A larger periodicity gives rise to an even richer adsorption scenario. The surface phase diagram of a model that is specified by $L^*=40$ and $L_1^*=35$ is depicted in Fig.~\ref{pd_L=40_L1=35}, where  an off-coexistence
extension of the bounded-wetting is present, in addition to groove-prewetting. This is illustrated in Fig.~\ref{L=40_L1=35_T127} for temperature $T^*=1.27$: the adsorption isotherm, which is shown in the upper panel, clearly
exhibits two first-order transitions that correspond to bounded-wetting and groove-prewetting. The loci of the equilibrium (stable) states are given by the concave envelope of $\Omega^{ex}=\Omega+pV$ displayed in the lower
panel; the precise locations of the two phase transitions can be determined as the cusps in $\Omega^{ex}(\mu)$.

The surface phase diagram of a substrate with somewhat broader grooves ($L^*=40$ and $L_1^*=30$) is further complemented by the groove-filling transitions, cf. Fig.~\ref{pd_L=40_L1=30}. For this model, the bounded-wetting
transition is unstable on the coexistence path ($T^*_{bw}=1.24$, $T^*_{gw}=1.23$). However, the curves that represent bounded-wetting and groove-prewetting intersect below the saturation line, which defines a triple point at
which three configurations of finite adsorption coexist.

For even broader grooves, the width of the ridges no longer permits bounded-wetting. For the model of $L^*=40$ and $L_1^*=20$ this results to the phase diagram where only groove-filling and groove-prewetting persist, as
illustrated in Fig.~\ref{pd_L=40_L1=20}. This behaviour is somewhat similar to the observation for the model with $L^*=15$ and $L_1^*=5$ (see Fig.~\ref{pd_L=15_L1=5}). However,  groove-filling is now stable with respect to
crystallisation up to $T_{gf}$, and its critical point extends even beyond $T_{gw}$.

\section{Discussion}

In this work,  phase transitions of simple fluids in a contact with grooved substrates have been investigated using a mean-field non-local density functional theory. The adsorption properties of such substrates were shown to
be remarkably complex as a consequence of an intricate interplay of various interfacial phenomena and sensitively dependent on the particular substrate geometry. The main conclusions that can be inferred form the presented
results are summarised as follows:

\begin{itemize}

\item When exposed to a saturated vapour, the wetting state of a grooved substrate may pass through four different regimes:

At lower temperatures, the system typically experiences the groove-filling transition, which is characterised by the condensation of the gas inside the grooves, provided the groove width is sufficiently large. For a given
periodicity, the value of $T_{gf}$ increases with the groove width. For narrow grooves, the groove filling, although taking place well above the bulk triple point, is already metastable with respect to wall-induced freezing.


The system becomes completely wet above a temperature $T_{gw}$, which is analogous to a wetting temperature $T_w$ for a flat wall. This groove-wetting transition is always first-order for the considered molecular model since
the range of the fluid-fluid and fluid-wall interactions is different; thus, their contributions to the Hamaker constant cannot be balanced.


Finally, our DFT study predicts the presence of another transition, which we call bounded-wetting transition. This transition (if stable) precedes groove-wetting and can be thought as a partial unbinding of a wetting film at
the ridges, the remnant of an ordinary wetting transition on a planar wall, which would occur at $T_w$ if the grooves were absent. Clearly, this transition is possible only if the ridges are sufficiently wide to accommodate
mesoscopically thick films. Some analogy can be found with thin-thick wetting on planar walls  for systems that exhibit long-range critical wetting \cite{lrcw1, lrcw2, lrcw3}. Unlike the latter, which is a consequence of the
competition of different interaction contributions, bounded-wetting is induced by lateral heterogeneity of the wall.

\item For all models that are considered here, we have found that $T_{gw}>T_w$, and in most cases, the difference between the two temperatures was significant.
This finding deserves some discussion considering that according to the macroscopic arguments, a corrugated surface of area $A_s$ becomes completely wet, if
 \bb
  \gamma_{sv}=\frac{A}{A_s}\gamma_{lv}+\gamma_{sl}\,, \label{anton}
  \ee
where $A<A_s$ is the area of the liquid-vapour interface. Because this condition is easier to be met than the one for a planar wall, one would expect that the wetting of a grooved surface occurs at a lower temperature than
wetting on a corresponding flat wall. Generally, macroscopic models such as the well-known Wenzel model \cite{wenzel} predict that the roughness of a solid surface enhances its wetting (or drying) properties; thus, corrugation
makes hydrophilic substrates even more wettable. Other simple phenomenological models, that consider the possibility of groove filling, provide qualitatively similar predictions. We claim that this contradiction can be
explained as follows:

1) Classical models rely on an analysis of a macroscopic liquid drop deposited on a solid surface whose contact angle is controlled by the surface tensions involved. However, a proper description of the models involving  tiny
capillaries \cite{comment} necessitates a more microscopic treatment \cite{rowlin}. In particular, the concept of the binding potential, which is neglected in macroscopic models, is particularly crucial \cite{evtar, rauscher}.
For the present molecular model the binding potential at the local height $\ell(x)$ decays as $\ell^{-2}(x)$ with an amplitude depending on the strength and geometry of the wall. As illustrated in Fig.~\ref{pot}, the effective
wall parameter is considerably weakened by the presence of the grooves, which tends to shift the temperature of complete wetting to higher values. Moreover, the molecular model presented here takes into account strongly
inhomogeneous character of the adsorbed fluid including packing effects that are particularly strong in the vicinity of the wall.

2) Our model of the grooved substrate exhibits sharp edges at the ends of the ridges. Here, the surface tension prevents the local meniscus to unbind from the substrate, even if far from the edge apex the height of the
interface above the surface is macroscopic \cite{edge}. Furthermore, the macroscopic Young equation (even if the line tension contribution is taken into account) ceases to hold if the three phase contact line is located at the
edge, since the apparent contact angle exhibits ambiguity. The latter phenomenon, known as Gibbs' criterion, has often been invoked in conjecture with contact line pinning at the edge, see e.g. Ref. \cite{yeomans}. Note
however, the concept of the contact line pinning was put questioned in the recent more microscopic study dealing with nanoscopic sessile droplets \cite{dutka} due to the presence of the accompanying wetting films.

\item Each phase transition that occurs at the two-phase coexistence also extends to the undersaturation states and terminates at its own critical point as showed
in Figs.~\ref{pd_L=15_L1=10}, \ref{pd_L=15_L1=5}, \ref{pd_L=40_L1=35}, \ref{pd_L=40_L1=30} and \ref{pd_L=40_L1=20}. In particular, the critical point corresponding to groove-filling occurs slightly above $T_w$ and the
difference is pronounced when the grooves (of a fixed depth) are wide (cf. Fig.~\ref{pd_L=40_L1=20}). This should be contrasted with the groove-filling in a deep ($D\to\infty$) grove, where, as mentioned in Section II, this
point corresponds to $T_w$ and separates first-order and critical filling regimes. We note that the limited numerical accuracy does not allow us to determine the way, at which the transition line connects the bulk coexistence,
so that this remains an open question.

\item In Ref. \cite{tasin}, the authors report their results based on effective interfacial Hamiltonian theory for complete wetting of patterned substrates (including rectangular grooves) with long ranged intermolecular
potentials and observed four distinct scaling regimes. This matches with the picture of adsorption for our model provided the isotherm corresponds to $T>T_{gw}$ and does not cross any of the first order transition lines. As an
example, we show in Fig.~\ref{final_plot} adsorption isotherm for one of the aforementioned substrate models at a sufficiently high temperature. Clearly, the filling regime corresponding to an abrupt rise of the meniscus
inside the grooves is less pronounced than that for much deeper capillaries considered in Ref. \cite{tasin}. The following postfilling regime is characterised by an almost linear dependence of the adsorption on
undersaturation. Finally, the last regime corresponds to the power law behaviour, Eq. ~(\ref{ell_pi}), as for the planar wall. The authors in Ref. \cite{tasin} further split the latter into the effective planar scaling regime
with a geometry-dependent Hamaker constant and the one, for which the geometrical patterns are irrelevant.

\begin{figure}[h]
\includegraphics[width=0.47\textwidth]{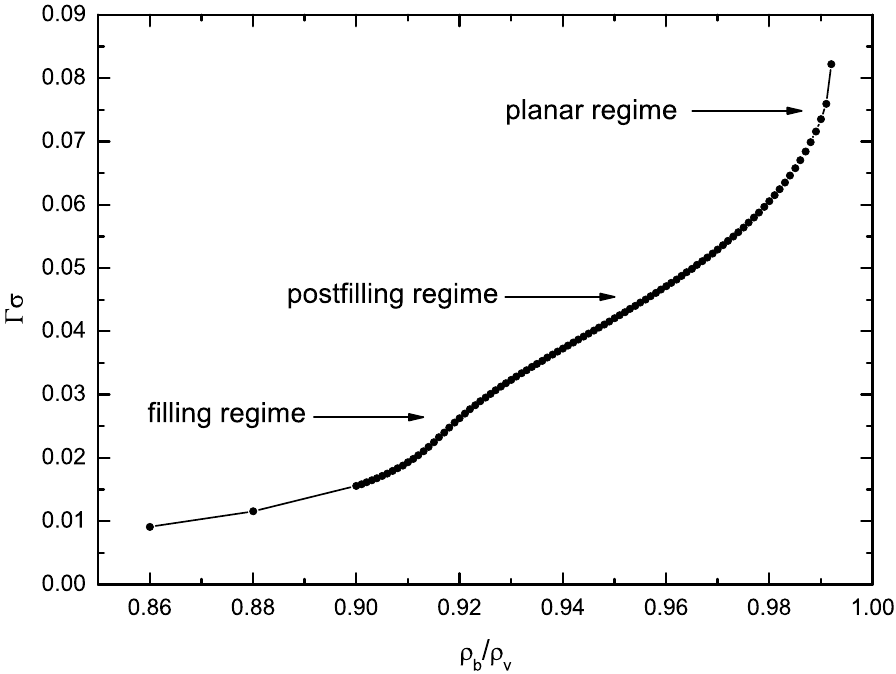}
\caption{Adsorption isotherm for a model with $L^*=40$, $L_1^*=20$, and $D^*=5$ at a temperature $T^*=1.35$. Filling, postfilling and planar regimes can be distinguished. The undersaturation is expressed as a ratio between the
bulk and saturated vapour densities.}\label{final_plot}
\end{figure}



\item The last remark concerns to the approximations and possible extensions of this work. As strictly of a mean field character, the approximative DFT necessarily
neglects some kinds of fluctuations such as those due to capillary waves and thus it is legitimate to ask whether the predicted phenomena could realistically be
observed. Recall, first, that since the molecular model includes dispersion forces, the mean-field approximation is valid for wetting and filling transitions in
three dimensions \cite{dietrich, lipowsky, rascon_05}. It is further important to emphasise that owing to an infinite extension of the groove array (together
with a translation invariancy along the $y$-axis), the phase transitions considered here terminate at true critical points. However, as already mentioned in
Section II, this would not be the case of a single groove (of finite depth), for which the groove filling transition must be rounded due to its pseudo-1D
character.

Throughout of this work the concern was on the fluid phase of the adsorbent only. As we have seen, however, a strong wall-induced crystallisation takes place well above the bulk triple point ($T_t^b\approx0.5\,T_c$, see, e.g.,
Ref.~\cite{tarazona_melt}). Therefore, at some cases (see Figs.~\ref{pd_L=15_L1=5} and \ref{pd_L=40_L1=30}) the crystal nuclei form before the groove-filling lines connect the bulk coexistence. One could avoid such issue by
considering lower value of the parameter $\varepsilon_w$ but then the other phase transitions would also be shifted and possibly vanish. The proper investigation of the equilibrium fluid solidification at the structured wall
is definitely an interesting task but requires a more demanding (3D) treatment than used here.

There is a number of open questions as regards to the related models. For instance, one may ask to what extent the phase behaviour scenario will be modified by considering the atomic corrugation of the wall, as addressed,
e.g., in Refs.~\cite{swain, rejmer_2000, kubalski}. In fact, the model itself could serve to mimic a periodically corrugated substrate if both $D$ and $L$ parameters are of the order of the molecular diameter $\sigma$. We note
that for $D\approx2\sigma$ the filling transition disappears. Also, it would be interesting to make a link with adsorption at chemically structured surfaces, especially in view of the ``morphological phase transitions''
predicted in Refs.~\cite{bauer, bauer2}.



\end{itemize}

\begin{acknowledgments}
 \noindent The support from the Czech Science Foundation, project 13-09914S, is acknowledged.
\end{acknowledgments}

\appendix

\section{Wall potential}

In this Appendix, the attractive contributions of the potentials $V_1(z)$ and $V_2(x,z)$ (Eqs.~(\ref{V1})--(\ref{V2})) as given by (\ref{V1_att})--(\ref{psi}) are derived.


We assume that the wall is formed with uniformly distributed atoms with a density $\rho_w$, which interact with the fluid particles according to (\ref{phiw}). A contribution of the semi-infinite planar wall, $V_1(z)$, is given
by integrating $\phi_w(r)$ over the domain $\{x\in\mathbb{R}, y\in\mathbb{R}, z\in(-\infty,0)\}$, which lead to a familiar $z^{-3}$ expression:

 \begin{eqnarray}
 \tilde{V_1}(z)&=&\rho_w\int_{-\infty}^\infty\dd x'\int_{-\infty}^\infty\dd y'\int_{-\infty}^0\dd z'\nonumber\\
 &&\phi_w\left(\sqrt{x'^2+y'^2+(z-z')^2}\right)\nonumber\\
 &=&-8\pi\varepsilon_w\rho_w\sigma^{6}\int_0^\infty\dd\rho\rho\int_{-\infty}^{-z}\dd z'\frac{1}{(\rho^2+z'^2)^3}\nonumber\\
 &=&\frac{2\alpha_w}{z^3}, (\forall z>0)\label{V1_apend}\,,
 \end{eqnarray}
 with $\alpha_w=-\frac{1}{3}\pi\varepsilon_w\rho_w\sigma^6$.

The attractive potential exerted by a single rectangular body comprising a volume $\{x\in(-L_1,0),y\in\mathbb{R},z\in(0,D)\}$ is:

 \begin{eqnarray}
 \tilde{V_2}(x,z)
 &=&\rho_w\int_{-L_1}^{0}\dd x'\int_{-\infty}^\infty\dd y'\int_{0}^{D}\dd z'\nonumber\\
 &&\phi_w\left(\sqrt{(x-x')^2+y'^2+(z-z')^2}\right)\nonumber\\
 &=&-4\varepsilon_w\rho_w\sigma_w^6\int_{-L_1}^{0}\dd x'\int_{-\infty}^\infty\dd y'\int_{0}^{D}\dd z'\nonumber\\
 &&\frac{1}{\left(\sqrt{(x-x')^2+y'^2+(z-z')^2}\right)}\nonumber\\
 &=&-\frac{3}{2}\pi\varepsilon_w\rho_w\sigma_w^6\int_{-L_1}^{0}\dd x'\int_{0}^{D}\dd z'\nonumber\\
 &&\frac{1}{\left[(x-x)'^2+(z-z')^2\right]^{5/2}}\nonumber\\
 &=&-\frac{3}{2}\pi\varepsilon_w\rho_w\sigma_w^6\int_x^{L_1+x}\dd x'\int_{z-D}^{z}\dd z'\frac{1}{\left(x'^2+z'^2\right)^{5/2}}\nonumber\\
 &=&-\frac{1}{2}\pi\varepsilon_w\rho_w\sigma_w^6\int_{x}^{L_1+x}\dd x' \left.\frac{z'(3x'^2+2z'^2)}{x'^4\left(x'^2+z'^2\right)^{3/2}}\right|_{z'=z-D}^{z'=z}\nonumber\\
 &=&\alpha_w\left[\psi_{z,D}(L_1+x)-\psi_{z,D}(x)\right]\,,
 \end{eqnarray}
 where
 \begin{eqnarray}
 \psi_{z,D}(x)&\equiv&\frac{2x^4+x^2(z-D)^2+2(z-D)^4}{2(z-D)^3x^3\sqrt{x^2+(z-D)^2}}\nonumber\\
&& -\frac{2x^4+x^2z^2+2z^4}{2x^3z^3\sqrt{x^2+z^2}} \,.\label{psi_apend}
\end{eqnarray}

It can be checked that for $z>D$:
 $$
 \lim_{x\rightarrow0}\psi_{z,D}(x)=\lim_{x\rightarrow\infty}\left[\frac{2(z-D)^4}{2(z-D)^4x^3}-\frac{1}{x^3}\right]=0
 $$
and
 $$
  \lim_{x\rightarrow\infty}\psi_{z,D}(x)=\frac{1}{(z-D)^3}-\frac{1}{z^3}\,.
 $$
 From these two limits it follows that  (using (\ref{V1_apend}))
 $$
  \lim_{L_1\to\infty}\tilde{V_2}(0,z>D)=\frac{\tilde{V_1}(z-D)}{2}-\frac{\tilde{V_1}(z)}{2}\,,
  $$
  as one expects.

Furthermore, for $z<D$:
$$
 \lim_{x\rightarrow0}\psi_{z,D}(x)=-\infty\,;
$$
thus, $ \tilde{V_2}(0,z<D)$ diverges as also expected.

Finally, in the limit of infinite $D$, (\ref{psi_apend}) becomes:

$$
\lim_{D\rightarrow\infty}\psi_{z,D}(x)=-\frac{1}{x^3}-\frac{2x^4+x^2z^2+2z^4}{2x^3z^3\sqrt{x^2+z^2}}
$$
and because
$$
\lim_{x\rightarrow\infty}\lim_{D\rightarrow\infty}\psi_{z,D}(x)=-\frac{1}{z^3}\,,
$$
one obtains in the limit of both $D\to\infty$ and $L_1\to\infty$
 \begin{eqnarray}
&&\lim_{L_1\rightarrow\infty}\lim_{D\rightarrow\infty}\left(\tilde{V_1}(z)+\tilde{V_2}(x,z)\right)\nonumber\\
&&=\alpha_w\left[\frac{1}{z^3} +\frac{2x^4+x^2z^2+2z^4}{2x^3z^3\sqrt{x^2+z^2}}+\frac{1}{x^3}\right]\,,
 \end{eqnarray}
which is the potential of a rectangular wedge, cf. \cite{malijevsky_prl}.

\section{Weighted densities}

In this Appendix, we derive the expressions of the Rosenfeld's weighted densities for a hard-sphere $\mathcal{N}$-component mixture, exploiting the rectangular symmetry induced by the external potential (\ref{pot}).

The weighted densities are defined by a sum of convolutions
 \bb
 n_\alpha(\rr)=\sum_{i=1}^{\mathcal{N}}\int\dr'\rho_i(\rr')w_\alpha^i(\rr-\rr')\,,
 \ee
 where, according to the original Rosenfeld fundamental measure theory \cite{ros}, the weighted functions consist of four scalars
 \ba
 w_3^i(\rr)&=&\Theta(R_i-r)\,;\;\;\;w_2^i(\rr)=\delta(\sigma/2-r)\,;\\
 w_1^i(\rr)&=&\frac{w_2(\rr)}{4\pi R_i}\,;\;\;\;\;\;\;\;\;\;\;\;w_0^i(\rr)=\frac{w_2(\rr)}{2\pi R_i}\,;
 \ea
 and two vectors
$$
{\bf w}_2^i(\rr)=\frac{\rr}{r}\delta(R_i-r)\;\;\;{\bf w}_1^i(\rr)=\frac{{\bf w}_2(\rr)}{4\pi R_i}\,.
$$


Here, $R_i$ refers to the radius of the $i$th species.

\subsection{$n_3(x,z)$}

The weighted density $n_3(x,z)$ is given by the following volume integral in the Cartesian coordinates:
\begin{eqnarray}
n_3(x,z)&=&2\sum_{i=1}^{\mathcal{N}}\int_{-R_i}^{R_i}\dd z'\int_{-\sqrt{R_i^2-z'^2}}^{\sqrt{R_i^2-z'^2}}\dd x'\nonumber \\
&&\sqrt{R_i^2-z'^2-x'^2}\rho(x+x',z+z')\,, \label{n3a}
\end{eqnarray}
which can be solved numerically by any standard quadrature. However, it may be more convenient to transform the dummy variables $z'\to R_iz'$ and $x'\to x' \sqrt{R_i^2-z'^2}$ to express $n_3(x,z)$ as the integral over the
domain $(-1,1)\times(-1,1)$:
\begin{eqnarray}
 n_3(x,z)&=&2\sum_{i=1}^{\mathcal{N}} R_i^3\int_{-1}^1\dd z'(1-z'^2)\int_{-1}^1\dd x'\label{n3}\\
 && \sqrt{1-x'^2}\rho(x+R_i\sqrt{1-z'^2}x,z+R_i z')\,.\nonumber
\end{eqnarray}
 This form allows us to employ the Gaussian quadrature:
 \begin{eqnarray}
 n_3(x,z)&\approx&2\sum_{i=1}^{\mathcal{N}} R_i^3\sum_k^{n_L}w_k^{L}(1-{(z_k^L)}^2)\\
 &&\left[\sum_j^{n_{c_2}}w_j^{c_2}\rho(x+R_i\sqrt{1-{(z_k^L)}^2}x_j^{c_2},z+R_iz_k^L)\right]\,,\nonumber
 \end{eqnarray}
 where $x_k^L$, $w_k^L$, $k=\{1,n_L\}$ are the nodes and the weights of the Legendre polynomials up to degree $n_L$,  and $x_k^{c_2}$,  $w_k^{c_2}$, $k=\{1,n_{c_2}\}$
 are the nodes and the weights of the Chebyshev polynomials of the second kind up to degree $n_{c_2}$.

\subsection{ $n_2(x,z)$, $n_1(x,z)$, $n_0(x,z)$}
The ``surface'' weighted function can be expressed as follows:
\begin{eqnarray}
n_2(x,z)&=&2\sum_{i=1}^{\mathcal{N}} R_i\int_{-R_i}^{R_i}\dd z'\int_{-\sqrt{R_i^2-z'^2}}^{\sqrt{R_i^2-z'^2}}\dd x' \nonumber\\
&&\frac{1}{\sqrt{R_i^2-z'^2-x'^2}}\rho(x+x',z+z')\,.\label{n2a}
\end{eqnarray}
Now, because the integrand blows up at the boundaries of the inner integral, the use of the Gaussian quadrature is vital. Following the same transformation as above, (\ref{n2a}) becomes
\begin{eqnarray}
n_2(x,z)&=&2\sum_{i=1}^{\mathcal{N}} R_i^2\int_{-1}^1\dd z'\int_{-1}^1\dd x'\\
&&\frac{1}{\sqrt{1-x'^2}}\rho(x+R_i\sqrt{1-z'^2}x',z+R_iz')\nonumber\\
&\approx&2\sum_{i=1}^{\mathcal{N}} R_i^2\sum_k^{n_L}w_k^{L}\label{n2}\nonumber \\
&\times&\left[\sum_j^{n_{c_1}}w_j^{c_1}\rho(x+R_i\sqrt{1-{(z_k^L)}^2}x_j^{c_1},z+R_iz_k^L)\right]\,,\nonumber
\end{eqnarray}
where $x_k^{c_2}$,  $w_k^{c_2}$, $k=\{1,n_{c_1}\}$  are the nodes and the weights of the Chebyshev polynomials of the first kind up to degree $n_{c_1}$.


Similarly, for $n_1(x,z)$ and $n_0(x,z)$ one obtains:
\begin{eqnarray}
n_1(x,z)&\approx&\frac{1}{2\pi}\sum_{i=1}^{\mathcal{N}}  R_i\sum_k^{n_L}w_k^{L}\\
&\times&\left[\sum_j^{n_{c_1}}w_j^{c_1}\rho(x+R_i\sqrt{1-{(z_k^L)}^2}x_j^{c_1},z+R_iz_k^L)\right]\nonumber
\end{eqnarray}
and
\begin{eqnarray}
n_0(x,z)&\approx&\frac{1}{2\pi}\sum_{i=1}^{\mathcal{N}}  \sum_k^{n_L}w_k^{L}\\
&\times&\left[\sum_j^{n_{c_1}}w_j^{c_1}\rho(x+R_i\sqrt{1-{(z_k^L)}^2}x_j^{c_1},z+R_iz_k^L)\right]\,.\nonumber
\end{eqnarray}

\subsection{${\bf n}_2(x,z)$, ${\bf n}_1(x,z)$}
The vectorial weighted densities can be dealt in the same manner. Thus, for the $x$-components of ${\bf n}_2(x,z)$ and ${\bf n}_1(x,z)$ we obtain
\begin{eqnarray}
n_2^x(x,z)&=&2\sum_{i=1}^{\mathcal{N}}\int_{-R_i}^{R_i}\dd z'\int_{-\sqrt{R_i^2-z'^2}}^{\sqrt{R_i^2-z'^2}}\dd x'\\
&& \frac{x'}{\sqrt{R_i^2-z'^2-x'^2}}\rho(x+x',z+z')\nonumber\\
&\approx&2\sum_{i=1}^{\mathcal{N}} R_i^2\sum_k^{n_{c_2}}w_k^{c_2}\nonumber\\
&&\left[\sum_j^{n_{c_1}}w_j^{c_1}x_j^{c_1}\rho(x+R_i\sqrt{1-{(z_k^{c_2})}^2}x_j^{c_1},z+R_iz_k^{c_2})\right]\,,\nonumber
\end{eqnarray}
and
\begin{eqnarray}
n^x_1(x,z)&\approx&\frac{1}{2\pi}\sum_{i=1}^{\mathcal{N}} R_i\sum_k^{n_{c_2}}w_k^{c_2}\\
&&\left[\sum_j^{n_{c_1}}w_j^{c_1}x_j^{c_1}\rho(x+R_i\sqrt{1-{(z_k^{c_2})}^2}x_j^{c_1},z+R_iz_k^{c_2})\right]\,.\nonumber
\end{eqnarray}


Similarly, for the $y$-components it follows that:
\begin{eqnarray}
n^y_2(x,z)&=&2\sum_{i=1}^{\mathcal{N}}\int_{-R_i}^{R_i}\dd z'\int_{-\sqrt{R_i^2-z'^2}}^{\sqrt{R_i^2-z'^2}}\dd x'\nonumber\\
&& \frac{1}{\sqrt{R_i^2-z'^2-x'^2}}\rho(x+x',z+z')\nonumber\\
&\approx&2\sum_{i=1}^{\mathcal{N}} R_i^2\sum_k^{n_L}w_k^Lz_k^L\\
&&\left[\sum_j^{n_{c_1}}w_j^{c_1}\rho(x+R_i\sqrt{1-{(z_k^L)}^2}x_j^{c_1},z+R_iz_k^L)\right]\,,\nonumber
\end{eqnarray}
and
\begin{eqnarray}
n^y_1(x,z)&\approx&\frac{1}{2\pi}\sum_{i=1}^{\mathcal{N}} R_i\sum_k^{n_L}w_k^Ly_k^L\\
&&\left[\sum_j^{n_{c_1}}w_j^{c_1}\rho(x+R_i\sqrt{1-{(z_k^L)}^2}x_j^{c_1},z+R_iz_k^L)\right]\nonumber
\end{eqnarray}

\section{$\int\dd\rr'\rho(\rr')u_{\rm a}(|\rr-\rr'|)$}

\begin{figure}[hbt]
\includegraphics[width=8cm]{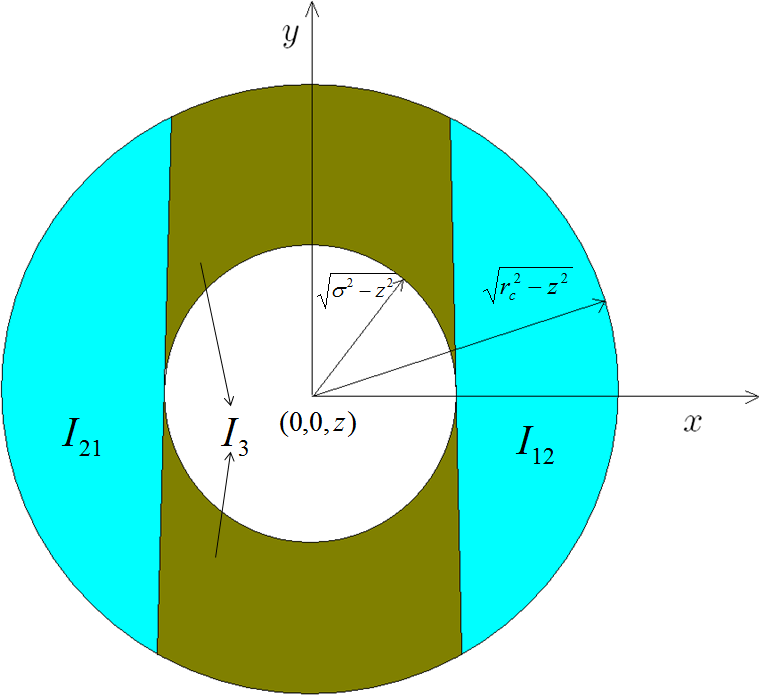}
\caption{Sketch of the integrated domain in the $x$-$y$ projection for $z<\sigma$.} \label{xy}
\end{figure}

In this Appendix, we carry out the third term in Eq.\,(\ref{el}):
 \bb
  I(\rr)\equiv\int\dr'\rho(\rr')u_{\rm a}(|\rr-\rr'|)=\int\dr'\rho(\rr+\rr')u_{\rm a}(r')\,,\label{I1}
\ee with the attractive potential $u_{\rm a}$ given by (\ref{ua}). With $\rhor=\rho(x,z)$, the function $I(\rr)=I(x,z)$ reads
 \begin{eqnarray}
  I(x,z)&=&\int u_{\rm a}(r')\rho(x+x',z+z')\dr'\label{I}\\
  &=&\int u_{\rm a}(r')\rho(x+x',z+z')\nonumber\\
  &&\times\left[H(r_c-r')-H(\sigma-r')\right]\dr'\,,\nonumber
\end{eqnarray}
where $H(x)$ is the Heaviside function, and we wish to express (\ref{I}) as a double integral over the $x$ and $z$ coordinates. Apparently, one can treat $I(x,z)$ as the difference between two independent contributions over
the spheres of radii $\sigma$ and $r_c$. Unfortunately, although this approach would be possible for a potential such as the square-well, it cannot be applied for the present model, since $u_{\rm a}(r)$ cannot be extended to
zero. Instead, $I(x,z)$ is separated into five terms as follows (see Figs.~\ref{xz} and \ref{xy}):

\begin{figure}[hbt]
\includegraphics[width=8cm]{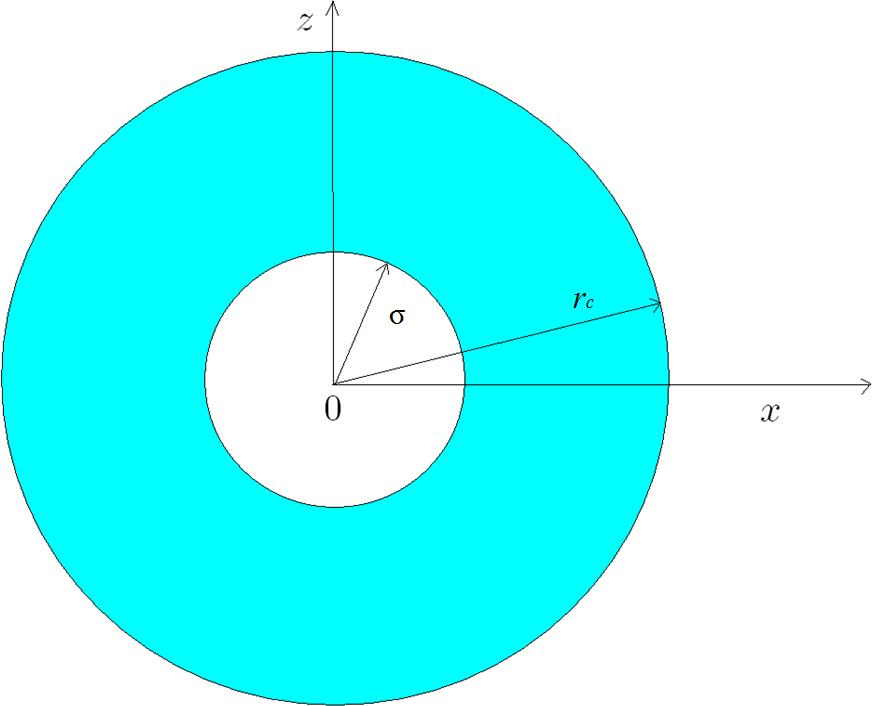}
\caption{Sketch of the integrated domain in the $x$-$z$ projection for $y=0$.} \label{xz}
\end{figure}

 \bb
 I=I_{11}+I_{12}+I_{21}+I_{22}+I_{3}\,,
 \ee
 where
 \begin{eqnarray*}
 I_{11}&=&-8\varepsilon\sigma^6\int_{\sigma}^{r_c}\dd z'\int_{-\sqrt{r_c^2-z'^2}}^{\sqrt{r_c^2-z'^2}}\dd
 x'\rho(x+x',z+z')\\
 &&\hspace{4cm}\times\psi_{r_c}(x',z')\,,
 \end{eqnarray*}
  \begin{eqnarray*}
 I_{12}&=&-8\varepsilon\sigma^6\int_{-r_c}^{-\sigma}\dd z'\int_{-\sqrt{r_c^2-z'^2}}^{\sqrt{r_c^2-z'^2}}\dd
 x'\rho(x+x',z+z')\\
 &&\hspace{4cm}\times\psi_{r_c}(x',z')\,,
 \end{eqnarray*}
  \begin{eqnarray*}
 I_{21} &=&-8\varepsilon\sigma^6\int_{-\sigma}^{\sigma}\dd z\int_{-\sqrt{r_c^2-z^2}}^{-\sqrt{\sigma^2-z^2}}\dd
 x\rho(x+x',z+z')\\
 &&\hspace{4cm}\times\psi_{r_c}(x',z')\,,
 \end{eqnarray*}
  \begin{eqnarray*}
 I_{22}&=&-8\varepsilon\sigma^6\int_{-\sigma}^{\sigma}\dd z'\int_{\sqrt{\sigma^2-z'^2}}^{\sqrt{r_c^2-z'^2}}\dd
 x'\rho(x+x',z+z')\\
 &&\hspace{4cm}\times\psi_{r_c}(x',z')\,,
 \end{eqnarray*}
and
   \begin{eqnarray*}
 I_3&=&-8\varepsilon\sigma^6\int_{-\sigma}^{\sigma}\dd z'\int_{-\sqrt{\sigma^2-z'^2}}^{\sqrt{\sigma^2-z'^2}}\dd
 x'\rho(x+x',z+z')\tilde{\psi}_{\sigma,r_c}(x',z')\,.
 \end{eqnarray*}
Here we use the following abbreviations:
   \begin{eqnarray*}
 \psi_{r_c}(x,z)&\equiv&\int_0^{\sqrt{r_c^2-z^2-x^2}}\frac{\dd y}{(x^2+y^2+z^2)^3}\\
 &=&\frac{5r^2y_2+3y_2^{3}}{8r^4r_c^4}+\frac{3\arctan\left[\frac{y_2}{r}\right]}{8r^5}\,,
 \end{eqnarray*}
  and
 \begin{eqnarray*}
 \tilde{\psi}_{\sigma,r_c}(x,z)&\equiv&\int_{y_1(x,z;\sigma)}^{y_2(x,z;r_c)}\frac{\dd z}{(x^2+z^2+z^2)^3}\\
  &=&\frac{1}{8r^5(r^2+y_1^2)^2(r^2+y_2^2)^2}\left\{r(y_1-y_2)\right.\\
  &&\left.\left[-5r^6+3y_1^3y_2^3+r^4(-3y_1^2+7y_1y_2-3y_2^2)\right.\right.\\
  &&\left.\left.+r^2y_1y_2(5y_1^2-y_1y_2+5y_2^2)\right]\right.\\
 &+&\left.3(r^2+y_1^2)^2(r^2+y_2^2)^2\left[\arctan\left(\frac{r}{y_1}\right)-\arctan\left(\frac{r}{y_2}\right)\right]\right\}\,,
\end{eqnarray*}
where $r(x,z)\equiv\sqrt{x^2+z^2}$, $y_1(x,z;\sigma)=\sqrt{\sigma^2-x^2-z^2}$, and $y_2(x,z;r_c)=\sqrt{r_c^2-x^2-z^2}$.

A singularity of $\tilde{\psi}_{\sigma,r_c}$ at $r\rightarrow0$ is removable:
$$
\lim_{r\rightarrow0}\tilde{\psi}_{\sigma,r_c}(x,z)=-\frac{1}{5}\left(\frac{1}{r_c^5}-\frac{1}{\sigma^5}\right)\,.
$$


\begin{thebibliography}{99}
\bibitem{comment}
Only shallow grooves were considered throughout of this work. No qualitative difference was found for the models with the deeper grooves (up to $D=20\sigma$).
\end{thebibliography}
\end{document}